# John Ellard Gore: of immensity and minuteness [1]

**Jeremy Shears**

## Abstract

John Ellard Gore FRAS, MRIA (1845-1910) was an Irish amateur astronomer and prolific author of popular astronomy books.  His main observational interest was variable stars, of which he discovered several, and he served as the first Director of the BAA Variable Star Section. He was also interested in binary stars, leading him to calculate orbital elements of many such systems. He demonstrated that the companion of Sirius, thought by many to be a dark body, was in fact self luminous. In doing so he provided the first indication of the immense density of what later became known as white dwarfs.

## Introduction

John Ellard Gore (1845-1910; Figure 1) was an Irish amateur astronomer and prolific author of popular astronomy books.  His main observational interest was variable stars. He was a founding member of the British Astronomical Association in 1890 and shortly after was appointed as the first Director of the Association's Variable Star Section (BAA VSS), which is the world's longest established organisation for the systematic observation of variable stars. Gore was a civil engineer by profession and served in India for 11 years before retiring to pursue his astronomical interests. Apart from variable stars, Gore was also intensely interested in binary stars, leading him to calculate the orbital elements of many such systems. In later years, most of Gore's attention was focussed on theoretical astronomy. He was the first person to estimate the density of a white dwarf (Sirius B), although, as we shall see, he rejected his result as he simply could not accept that matter of such high density could exist.

In this paper I describe Gore's life and his contributions to astronomy, drawing on his own writings, including his papers, his letters to the *English Mechanic and World of Science* (2) and his books. I also referred extensively to three obituaries, for the *Monthly Notices* of the RAS (3), *The Observatory* (4) and *Popular Astronomy* (5), as well as the biographies written by A.P. Fitzgerald in 1964, publish posthumously (6), by Charles Mollan (7) and by Ian Elliott in *The Biographical Encyclopaedia of Astronomers* (8). Surprisingly, Gore was afforded only a very brief obituary in JBAA (9) and today he is known by rather few members of the Association.

## Family and career

Gore was born in Athlone, Ireland, on 1 June 1845. He was the eldest son of the Venerable John Ribton Gore (1820-1894), Archdeacon of Achonry, County Sligo, also Canon of Killala and Rector of Dromard, and Frances Brabazon Ellard (1816-1896), from whom he obtained his second name, stylising himself J. Ellard Gore (7). The Gore family were descended from Sir Paul Gore (1567-1629), who was born in London and went to Ireland in 1602 as a commander of a troop of horse. In 1615 he became MP for Ballyshannon also in County Donegal (10) and in 1622 he was created Baronet of Magherabegg, in the County of Donegal (11). The Gore family seat was Manor Gore in Donegal. John Ellard Gore's great grandfather, Arthur Saunders Gore (1734-1809), was the Second Earl of Arran and the Arran Islands.

John Ellard Gore had 3 brothers (12) and a sister (13). He was educated privately and went on to Trinity College Dublin, receiving a Diploma in Civil Engineering in 1865 (14). After working as a railway engineer in Ireland for over 2 years, he passed second in the India Office open examination of candidates seeking appointments in the Indian Public Works Department (15). As a result he was appointed Assistant Engineer on the Sirhind Canal project (16) in the Punjab. He remained in this position until 1877 at which point he returned to Ireland on two years leave of absence. Apparently he did not get on well with his boss (17) and at one time wrote to an Indian newspaper criticising the policy of the Department. This was probably the reason why in 1879, whilst still on leave, he decided to retire from the Department after only 11 years service with a pension in hand. At this time he resided at Ballisodare, County Sligo, with his father until the latter died in 1894, at which point he moved to lodgings in Dublin where he spent the rest of his life. He never married and devoted his life to astronomy, writing papers and books on astronomy. During the mid 1880s he endeavoured to find employment both as an engineer and as an astronomer, but without success (18).

**An interest in astronomy**

It is not know what stimulated Gore's interest in astronomy. It has been suggested (19) that since Gore's father's house was only about 10 miles from Edward Joshua Cooper's (1798-1863) observatory at Markree Castle, County Sligo, it is possible that the young Gore visited the observatory and his interest began there. Cooper (Figure 2) had a 13½ inch (35.5 cm) refractor which he used up to his death in 1863. The observatory subsequently fell into disuse until 1874 when August William Doberck (1852-1941) was appointed Director, serving until 1883. He was succeeded by Albert Marth (1828-1897) who was Director until 1898. The observatory was used intermittently until 1902 at which point the telescope was sent to the Hong Kong Observatory. However, there is no direct evidence that Gore met Cooper, Doberck or Marth at Markree. Gore might have met Doberck through the Royal Irish Academy, whose headquarters are in Dublin, as both were Members (20).

Whenever they actually began, Gore's astronomical pursuits really took off during his time in the Punjab. As well as using his naked eyes and binoculars, he had a 3 inch (7.5 cm) f/16 refractor by Browning of London (21) and a 3.9 inch (10 cm) f/15 refractor by Wray, which was on an equatorial mount by Clarkson. His first book, *Southern Stellar Objects for Small Telescopes* was based on observations made using these instruments whilst in the Punjab and was published in India in 1877 (22) .

However, Gore found the summer months in India were not particularly conducive to observing the night sky: (23)

"In the Punjab summer, when the air is extremely dry and hot (sometimes over 100° F [38° C] several hours after sunset), it is generally so laden with dust that telescopic observation is out of the question—stars below 4 mag. being invisible to the naked eye". (24)

The British Raj had a number of hill stations in the foothills of Himalayas that allowed people to escape the heat of India's cities during summer. Not only were such elevated places cooler, but they offered clear skies comparatively free from the haze and dust of lower lying regions. Gore was particularly impressed with the naked eye views of the Milky Way from the hill station at Kussowlee. In September 1872 he drew attention to (25):

"a remarkable vacuity in or near the Milky Way in Cygnus. This dark space has for some time attracted my notice, as in all the star maps in my possession.........the Milky Way is shown continuous over the point in question. If a line be drawn from γ Cygni through α (Deneb) it will pass through this vacant place, at about the same distance from α that α is from γ. The small double star marked [Piazzi] 429 is near the southern boundary of this "coal sack," from which a dark rift passes across the Milky Way between the stars ζ and ρ Cygni. These remarks, of course, only refer to naked eye observation, as even with a very small telescope I have seen numerous small stars scattered over the space alluded to. In the clear air of the Himalayas, at an altitude of 6,000 feet [1800 m] above the sea, this dark spot is particularly noticeable, and immediately attracts the eye when directed to that portion of the heavens".

A modern image of the Milky Way in the Cygnus-Cepheus region appears as Figure 3 and this shows the void. The following year he also observed "a very faint offshoot of the Milky Way" in the region of $\pi^1$ to $\pi^6$ Orionis (26).

**Variable star observations and discoveries**

There is no record of precisely when Gore made his first variable star observation. He clearly had an interest in such objects in 1872 as in that year he wrote a letter (27) to the *English Mechanic* describing some of the main types of variables, although he did not indicate whether at that stage he was actually observing them. The following year, letters began to appear reporting his magnitude estimates of various stars. He had in his possession whilst in India copies of several star atlases and he began to notice that the stars plotted didn't always match reality (28). For example (29):

"On the 13th July [1874], while looking at Scorpio, I observed a star (apparently) of about 4th mag., a little north, about 1°, of $\zeta^1$ and $\zeta^2$ Scorpii. There is no star marked in this place in Mr. Proctor's Atlas, but in Keith Johnston's Atlas of Astronomy I see there is one shown in or near this place, and marked "variable." On looking at the object with the telescope I was surprised to find that it was really a beautiful cluster of stars, three of which are nearly equal, about 6th or 7th mag., and the remainder fainter. It is situated about midway between ζ and three 6th mag. stars, forming a triangle (between ζ and μ). Can any of your astronomical readers tell me anything about this variable and cluster?"

This chance observation stimulated a more thorough survey of other stars that had been suspected by earlier observers to be variable, many of which were subjects of letters he wrote to the *English Mechanic*. Between 1874 and 1877 he made estimates of 48 stars which he described in *Southern Stellar Objects for Small Telescopes* (22). In 1884 he presented the Royal Irish Academy with a *Catalogue of Known Variable Stars with Notes and Observations*, listing 191 such objects (30). The following year he published a *Catalogue of Suspected Variable Stars* (31) and in 1888 he issued a *Revised Catalogue of Variable Stars* with 243 objects, reflecting the rapid pace of discovery of such objects. He sent copies of the first *Catalogue* to a number of well-known astronomers, including Captain William Noble (1828-1904), who later became the first President of the BAA, Isaac Roberts (1829-1904), and E.B. Knobel (1841-1930), President of the RAS and the prominent variable star observers Joseph Baxendell (1815-1887) and George Knott (1835-1894; Figure 4). It was about this time that Gore was investigating possibilities for employment as an astronomer, so it might have served his purpose to ensure that his work was more widely known.

In the course of his systematic observations, Gore discovered several variable stars including EW CMa (32), W Cyg, S Sge, U Ori and X Her. He announced his discovery of W Cyg in a brief note in *Astronomische Nachrichten* (33), dated 9 September 1885:

"I beg to announce the discovery of variability in the red star No. 587 of Birmingham's Catalogue of Red Stars. The star lies a few minutes of arc s[outh] f[ollowing] ρ Cygni......The variation seems to be from about 5.8 mag to 7.5 mag with a period of 250-300 days, but further observations will be required to determine the period exactly. The star was at a maximum at the end of December 1884, at a minimum early in June 1885, and at present it is not far from maximum again".

W Cyg is now classified as a semi-regular variable of the type SRb. An analysis of 89 years of BAA VSS observations by Howarth (34) has revealed two independent periodic variations of approximately 131 and 234 days. In 1889 Gore sent a complete list of his observations of W Cyg to Paul S. Yendell (1844-1918) of Dorchester, Massachusetts, USA (35), including a list of maxima and minima which is shown in Figure 5. Yendell, a veteran of the American Civil War, was one of the most active American variable star observers of the period, accumulating more than 30,000 observations (36). Gore communicated regularly with him and enlisted his assistance in obtaining follow-up observations of X Her after Gore discovered it (37).

In October 1885 Gore announced his discovery of S Sge, which he had first suspected of variability back in 1879 (38). He reported a short period of 8-9 days and his observations were confirmed by E.F. Sawyer (1849-1937) in the USA and Rev. Thomas Henry Espinell Compton Espin (1858-1934; Figure 6) in Great Britain. It is now known to be a Cepheid variable with a period of 8.4 days and a range from magnitude 5.2 to 6.0.

Then on the evening of 13 December 1885 Gore detected what was initially thought to be a 6[th] magnitude Nova in Orion (39):

"I first saw this star with a binocular field-glass.....at 9.20 P.M., Dublin mean time. My attention was attracted to it by its very reddish colour and its absence from Harding's charts"

He informed Ralph Copeland (1837-1905; Figure 7) at Dun Echt Observatory, Aberdeen, of his discovery on 16 December. Copeland confirmed it the same evening (40) and sent a telegram (41) of the news to Harvard College Observatory (HCO), USA, as well as distributing a circular to other astronomers around the world (42). HCO picked the object up later the same night at magnitude 6.0. Spectroscopy at HCO and the Royal Greenwich Observatory (43) showed that it was not a nova, but a Mira-type Long Period Variable. The star is now known as U Ori (44) and is the first Long Period Variable to be identified via a photograph of its spectrum. Gore's analysis of observations during the first 5 years since discovery yielded a period of 373.5 days (45). Analysis of data in the BAA VSS between 1886 and 2006 yields an average period of 371.8 days and a typical range of magnitude 6.5 to 12.0, although an extreme maximum of magnitude 5.0 was observed in 1973 and an extreme minimum of magnitude 13.3 in 2006 (46). A light curve during the interval 1914 – 1918 is shown in Figure 8. Recently U Ori has become the first Long Period Variable to be shown to be asymmetrical via infra-red observations made during a lunar eclipse (47).

**Director of the BAA Variable Star Section**

Gore was elected FRAS on 8 March 1878 whilst resident in the Punjab. The signatories to his RAS Fellowship proposal form (Figure 9) are a veritable who's who of nineteenth century astronomy. His proposer "from personal knowledge" was the London instrument maker John Browning (1835-1925; Figure 10), who had supplied Gore's 3 inch refractor. The two corresponded regularly during Gore's time in the Punjab and on occasion Browning passed on his observations to scientific publications, such as those of the 1864 transit of Venus (48). The other signatories were: William Huggins (1824 – 1910; Figure 11), pioneer astronomical spectroscopist, William Lassell (1799-1880; Figure 12), renowned amateur astronomer and discoverer of Neptune's moon Triton and Uranus's moons Ariel and Umbriel, Edmund Neison (49) (1849-1940; Figure 13), selenographer, and Herbert Sadler (1856-1898) (50).

As mentioned earlier Gore was a Member of the Royal Irish Academy and served for a time on its Council. He also belonged to the Royal Dublin Society whose mission was to encourage the development of science, arts and industry throughout Ireland. Gore was a member of two overseas astronomical societies: the Société Astronomique de France and the Astronomical and Philosophical Society of Toronto, Canada (51). He was also an active member of the Liverpool Astronomical Society (LAS), which was founded in 1881, serving as its Vice President for a time (52). Despite its name, the LAS had a national membership and an international reputation; in many ways it was the precursor of the BAA. The LAS had a variable star section which was formed in 1884 (53) with Espin (Figure 6) as Director (54). Later the same year Gore became its second Director, remaining in post until 1889, when the LAS in its current form essentially collapsed. The BAA was formed in October 1890 (55) and many LAS members joined immediately, Gore being an inaugural member (56).

The BAA was organised along the lines of observing sections and it was natural that Gore was invited to become the first Director of the VSS, given his reputation in the field and his experience as Section Director at the LAS. At the meeting of 31 October 1890 he proposed to the Council the following programme for the new BAA VSS: (57)

- The observation of known variable stars which have for some cause been neglected by variable star observers
- The observation of stars suspected on good grounds to be variable, and of suspicious objects met with in the course of the observations
- A systematic search for variables of short period in selected portions of the short period zone (Pickering's)
- The supply to members of comparison stars and information respecting variable stars generally

In 1892 he proposed an addition to the programme (58): a nova search plan. This was facilitated by supplying charts to prospective observers showing stars to magnitude 6.7 for selected regions of the Milky Way.

Gore's role as Director of the BAA VSS (and LAS VSS) is described in more detail in an excellent paper on *British Variable Star Associations, 1848-1908* by John Toone (59). Although Gore published reports of members' observations in the BAA *Journal* and *Memoirs*, these were generally organised along the lines of reports of individual observers' data, just as they had been at the LAS. In fact at the time it was common for observers to publish reports of their own observations. It wasn't until later in his Directorship that Gore combined data from different observers in order to indentify maxima and minima of the Long Period Variables under observation. What he did not do was to facilitate cooperation amongst observers on a limited programme of stars. At the same time, the focus of the Section was increasingly on the nova search programme and observations of suspected variables, which by their nature generally yield negative results (and a few false alarms in the case of the nova programme). As Toone points out (59), this was a strategic mistake as many variable star observers, especially new ones, like actually to *see* the variable star and *detect* obvious variations. Moreover, Gore lacked the leadership that was required to move the Section forward and as the 1890s advanced membership was slipping. Without the feedback that their results were being made use of, fewer members were sending in reports with some preferring to submit their data elsewhere. Gore's Director's report for 1897 indicated the low ebb that the Section had reached: (60)

"The Director has very little to report and simply desires to state that the work of the Section has been continued during the past session on the same lines as in former years. Known and suspected variable stars have been regularly observed and the search for new or temporary stars along the course of the Milky Way continued"

Gore retired as BAA VSS director in 1899 and was succeeded by E.E. Markwick (1853 – 1925; Figure 14) (61). Markwick was a serving army officer and his military background brought with it the leadership and organisational skills required to reinvigorate the VSS (62). He set about organising its work along lines that are largely pursued to this day and which other variable star organisations around the world have emulated. Markwick, on behalf of BAA VSS members and others, arranged for Gore to be presented with a book of photographs taken by Isaac Roberts "in recognition of his services as Director of the Section during a period of nine years, 1890-99, in which each member has received from him much valuable help and encouragement in the observation of Variable Stars" (63). Whilst Gore might have been much respected by the members, Markwick became much loved. He was gregarious, affable and through his letters to individuals and frequent circulars advising them of the latest happenings, he provided the necessary feedback and encouragement to stimulate them to make further observations.

**A New Star in Perseus**

One of the most remarkable astronomical events of Gore's life occurred in 1901. In *Studies in Astronomy* he records (64):

"On the evening of Friday, February 22, 1901, while returning home from the house of a friend in Dublin, about 11h 40m P.M., Greenwich Mean Time, I happened to look towards the constellation Perseus, and was astonished to see a bright star of nearly the 1st magnitude shining in a spot where I knew that no star visible to the naked eye had previously existed. Next morning I telegraphed to the Observatories at Greenwich and Edinburgh, and also to Sir William Huggins, the famous astronomer".

Gore had thus made an independent discovery of Nova Persei 1901. However, he learnt from a reply from Ralph Copeland, at that time Astronomer Royal for Scotland, that he was not the first to see the Nova. In the early hours of the morning of 22 February, at 02:40 UT, the Edinburgh amateur astronomer Dr. Thomas David Anderson (1853-1932) was about to retire for the night when he cast a final "casual" glance up to the sky and noticed the third magnitude star in Perseus (65). The object was also seen earlier in the evening of 22 February, at 18:40 UT, by Ivo F.H.C. Gregg of St. Leonards, Sussex. Gregg, not being certain what the object was, sent a telegram to BAA VSS Director, E.E. Markwick, who confirmed it (66).

Nova Per 1901 (now known as GK Per) was one of the brightest novae of modern times, reaching magnitude 0.2 at its peak. Gore and other members of the BAA VSS, as well as astronomers around the world, monitored it intensively.

Gore had maintained a strong interest in novae from the time he observed his first one, Nova Cygni 1876 (Q Cyg) (67). And as we saw earlier, in 1885 for a short time he thought he had discovered his own nova in Orion, but which turned out to be a Long Period Variable. He was keenly interested in the "Temporary Star" of 1604, "Kepler's star" in Ophiuchus, which is now known to have been a supernova, writing about it several times over the years. He even provided a finder chart of its location in his *Southern Stellar Objects for Small Telescopes* (Figure 15) should readers wish to monitor the field for a possible recurrence. In October 1874 he thought he might have stumbled across this object brightening again: (68)

"Very close to the position assigned to the great new star of 1604, in R.A. 17h. 23m. S. 21º 16' (about midway between ζ and 58), I see a star of about 9$^{th}$, or perhaps 10$^{th}$ mag., not marked in Harding's large " Atlas Novus Cœlestis" (1822), just received (in which stars to 10 mag. are shown). This may possibly be the Temporary Star again brightening".

But, of course, it was not to be and we now know that supernovae do not recur (69). Gore also invoked a recurrence of Kepler's Star as a possible explanation of the "Broughty Ferry Mystery". Richard Baum describes this incident in his book *The Haunted Observatory* (70), but the essence is as follows. On the morning of 21 December 1882, several people at Broughty Ferry, near Dundee, noticed a bright star-like object "in close proximity to" the Sun during daylight. A report appeared in the *Dundee Advertiser* (71) the following day and three days later an anonymous writer (72) suggested it was probably the planet Venus. R.A. Procter (1837-1888) picked the story up in the first issue of *Knowledge* for 1883 (73), noting that at the time of the observation Venus was simply too far away from the Sun to be a valid explanation. Gore, in the same edition, also dismissed Venus as the explanation for the same reason: (74)

"With reference to the phenomenon.....if the object was not a comet near perihelion, it was probably a temporary star. I find there is no bright star near the place, the nearest being ε and σ Sagittarii (of the 2$^{nd}$ mag.), each about 12º distant; but these would be south of the sun, and not "a little above the sun's path", as the description states, besides being too faint to be visible in the daytime (to the naked eye) under any circumstances. The object could not have been the planet Venus, which was situated about 23º west of the sun on the day in question. It seems worthy to remark that the place of Kepler's celebrated "Nova", of 1604, was – at the time of

observation – only about 8½° to the west (and a little north) of the sun's place. So that, if the object was really a star, it seems possible that it may have been another outburst of Kepler's star".

Having developed the hypothesis, Gore went on to suggest (74) that "The morning sky should be examined, as if the object be still visible, of the same brilliancy, it should now be a conspicuous object before sunrise".

Following his analysis of the known facts, Baum proposes (70) that, contrary to Gore's and others' assertions, the most likely explanation of the mystery object was indeed a daylight observation of Venus. He notes that the phrase "*in close proximity to the sun*", when used by a layman, is a *concept* and not a *measure*. Thus to the laymen at Broughty Ferry, an angular separation of 23° could be described as being "in close proximity".

**Binary stars and white dwarfs**

Apart from variable stars, Gore's other astronomical passion was double stars. Although he enjoyed observing these objects with his telescopes, his main interest was computing the orbits of binary systems. He published the orbital elements of many binaries in a variety of journals, starting with γ Corona Australia in 1886 (75). After this the floodgates opened and 5 more papers followed the same year (40 Eri (76), ζ Sgr (77), OΣ 234 (78), τ Cyg (79) and Σ 1757 (80)) with a further 6 the following year. In 1890 he presented to the Royal Irish Academy a catalogue of 59 binary stars (81).

It was during his analysis of binary stars orbits that he became the first person to estimate the mass of a white dwarf. The process which led to his conclusion is a good example of the methodical approach he of took to many astronomical matters, starting with known facts and some basic calculations, then extrapolating in small, logical, steps to a solution (82). This sets him apart from many amateur astronomers of the time who were simply content to observe and record what they saw. Gore did indeed observe things, but he wanted a mathematical understanding of them too.

Sirius had been suspected of being a binary system since 1844 when Wilhelm Friedrich Bessel (1784-1846) proposed an unseen companion as being responsible for perturbations in its proper motion (83). The companion was finally revealed on the evening of 31 January 1862 when Alvan Graham Clark, testing an 18½ inch lens at the Clark home in Boston, Massachusetts, turned it towards Sirius (84). The companion's presence was immediately confirmed by his father, Alvan Clark. Sirius B, as the companion is now called, was some 10 magnitudes fainter than the primary which led to questions about whether it was this object that was actually causing the perturbations, or whether perhaps it was a close field star. There were also suggestions that Sirius B was not actually a star, i.e. a self-luminous object, but rather a large planet reflecting the light of Sirius. However, Otto Wilhelm Struve (1819-1905) demonstrated the gravitational connection between the two objects in 1864 (85) and in 1866 he estimated the mass ratio of the two stars to be 2.09:1 (compared to the modern value of 2.03:1). He drew attention to the huge difference in luminosity of the two objects, concluding that they "are of very different physical constitution" (86).

As is so often the case, when one knows of an object's existence, it becomes much easier to see, especially when one knows where to look. Thus after news of Clark's observation of Sirius B spread, it was widely observed by others with much smaller instruments. One such was the prolific variable star observer George Knott (Figure 4) from his observatory at Cuckfield, Sussex. Knott was the first to comment on the colour of the companion: (87)

"Happening to turn my 7 1/3 inch Alvan Clarke [*sic*] refractor on Sirius, on the 24th of January [1866], I was surprised to find the small companion, not withstanding bright moonlight, a tolerably conspicuous object. Its colour was a fine pale blue (about Blue$^3$ of the late Admiral Smyth's chromatic scale)"

Given Gore's interest in binary stars and their orbits, it was natural that he should turn his attention to the Sirius system. In 1889 he computed the elements using observations published since the discovery of the companion (88). Having performed the standard orbital calculations, which by then were routine to him, he addressed square on the controversy about whether or not the companion is self-luminous. In an 1891 JBAA paper he wrote: (89)

"It has been suggested, I believe by [William Rutter] Dawes, that the companion of Sirius, discovered by Alvan Clark, may possibly shine merely by reflected light from its primary. To test the probability of this hypothesis I have made some calculations, and although the assumed data are of course necessarily somewhat uncertain, the result I have arrived at will, I think, show that such hypothesis is quite untenable".

Using existing parallax measurements of Sirius, he estimated its distance and compared its apparent magnitude with that of the Sun, concluding that Sirius is "about 40 times brighter than the Sun would be at the same position". He then calculated the mean distance between the primary and companion as 22 times the Sun-Earth mean distance. Making various assumptions about the size, distance and albedo of the companion he estimated that, if it were purely reflecting the light of the primary, it would appear as an object of magnitude 16.6. In case the significance of this result had escaped anyone's attention, he went on to point out: "A star of this magnitude so close to Sirius would of course be quite invisible in our largest telescopes......It seems clear therefore that the companion must shine with some inherent light of its own, otherwise it could not possibly be so bright as the 10th magnitude, the brightness at which it is generally estimated" (89). Based on these results, he later concluded (90), as had Struve before him, that "The two bodies must be differently constituted".

He revisited the problem of Sirius' faint companion in 1905, noting (91) that "the faintness of this star must be due either to its small size or to its small luminosity of surface per unit area". Pursuing the former hypothesis, and again making a number of reasoned assumptions, he estimated that the companion would have a density 44,282 times that of water. He immediately rejected out of hand such an apparent absurdly high density, stating that "This is, of course, entirely out of the question". Instead he preferred the alternative explanation of a large low luminosity object, possibly "a very large globe of very low density and small luminosity, like the gaseous nebulae". He very much hoped that in future a spectroscope might reveal the true identity of the companion.

We now know that Sirius B is a highly evolved star known as a white dwarf, with a density even greater than Gore's rejected estimate. Current data suggests it is a tiny object with a radius of 0.0084 that of the Sun (5840 km) and a mean density of 2.38 x $10^6$ g/cm$^3$, more than 50 times Gore's estimate.  At the time that Gore made his calculation it was simply inconceivable that matter could exist at the density he calculated, so it was natural for him to reject his estimate, but he was clearly on the right track. Gore was also the first to comment on the unusual faintness of the companions of 40 Eridani (81) and Procyon, suggesting that the latter object might not be a star at all, but a low density gaseous nebular (91). Both companion stars, 40 Eridani B and Procyon B, are now also known to be white dwarfs.

As was the case with Sirius B, it had also been suggested that Algol's companion, might be a dark object.  In 1880, the Harvard astronomer Edward Charles Pickering (1846-1919) presented evidence that Algol was an eclipsing binary (92). This was confirmed in 1889, when Hermann Carl Vogel (1841-1907) at Potsdam found periodic Doppler shifts in the spectrum of Algol, inferring variations in the radial velocity of this binary system (93). Gore used a similar line of reasoning as he applied in the case of Sirius B to investigate the matter, concluding (94) "if the term "dark" is meant an object resembling the earth or even Jupiter, I fail to see there is any warrant for such a suggestion". Current understanding is that Algol A is a main sequence star, while the less massive Algol B is a sub-giant star at a later evolutionary stage.

**Giant and Miniature Suns** [95]

One of Gore's closest friends was the amateur astronomer William Henry Stanley Monck (1839-1915; Figure 16), who lived only a few streets away from him in Dublin (96). It was Monck's letter to the *English Mechanic* in July 1890, suggesting that a national amateur astronomical association with headquarters in London be set up, that acted as a catalyst that led to the formation of the BAA later that year (55). Like Gore, Monck was a graduate of Trinity College Dublin. He was called to the Bar in 1873 and he served as Chief Registrar for the Bankruptcy Division of the High Court of Ireland. In 1878 he returned to academic life as Professor of Moral Philosophy at Trinity. He wrote several works on logic, metaphysics and astronomy. A keen amateur astronomer, he had a fine 7½ inch refractor with an objective by Alvan Clark; the lens had originally been used in one of W.R. Dawes's telescopes (97). This telescope was used in September 1892 in the first experiments to measure starlight electrically (98). Monck's friend George M. Minchin (1845-1914), himself a Trinity graduate, but at the time living in London (99), had developed a photo-voltaic cell, comprising a selenium photo-cathode on an aluminium substrate immersed in acetone, that he wished to try out on Monck's telescope. During August 1892 progress was thwarted by clouds and Minchin eventually returned to London. Then on the morning of 28 September Monck, assisted by his neighbour, Stephen M. Dixon (1866-1940) (100), succeeded in measuring the relative brightness of Jupiter and Venus. Whether Gore was involved in some way is not known, but he must at least have been aware of the experiments. Minchin went on to make further successful photoelectric observations of stars in 1895 and 1896 using the 24 inch reflector at Daramona Observatory in County Westmeath (101).

In 1887 Monck had developed an equation relating the relative brightness of two stars in a binary system to the orbital period, their orbital separation and their apparent photometric intensity (102). He expressed the brightness relative to a reference star, ζ UMa. Gore used this equation to calculate the relative brightness of the 59 binaries in his catalogue (81), resulting in a huge range of relative brightness from 93 to 0.0015 times that of ζ UMa, with the faintest being 40 Eridani B.

In addition to addressing the large apparent differences in brightness of stars within binary systems, Gore also investigated the intrinsic luminosity of other stars and concluded that different stars are likely to have different sizes. Once again, Gore used reasoned assumptions and simple calculations, based on parallax and brightness, to assess the relative size of various stars (103). In the case of Arcturus, he concluded it was a "giant star" about 100 times the diameter of the Sun. Similarly he estimated that Capella was a "giant" with a diameter 18 times that of the Sun. Present day measurements place the diameters of Arcturus and Capella at 25.7 and 12.2 solar diameters respectively. Gore considered other stars, but in many cases simply gave a qualitative assessment of their size. Other "giant suns" included α Centauri, Deneb, β UMi, Canopus, β Cen, whereas Sirius could not "be considered a 'giant sun'". He found Vega to be "large", but smaller than Arcturus; Rigel was "enormous" and Pollux "seems to be largest of all".

At the other end of the scale, Gore suggested "it seems highly probable that there are miniature as well as giant suns". Noting that the stars in the Pleiades cluster are presumably at a similar distance from Earth, the range in brightness suggests a diameter range of 631 and some might be as small as Jupiter.

In 1894 Monck tackled the problem from a slightly different angle, taking into account their spectral colour. In his paper on "The spectra and colours of stars", published in this *Journal* in 1894 (104), he concluded there were probably two distinct classes of yellow stars, "one being dull and near to us and the other bright and remote". This was again a clue to the existence of what are now known as giant and dwarf stars. Gore took the analysis further in 1905 and listed spectral type and relative brightness for a variety of stars. Had he known it, he had to hand in the paper sufficient data to have constructed what would later become known as the Hertzsprung-Russell diagram, several months before Ejnar Hertzsprung's (1873-1967) first attempt.

## The structure of the universe and the brightness of the sky

Gore was fascinated by the form of the Milky Way and the fact that there were various apparently star-less regions such as the one in Cygnus that he observed whilst in the Punjab. This observation led him to propose a star count project in a letter to the English Mechanic: (105).

"counting all stars in the same field of view in the first instance with the brightest stars, and secondly, with those of the 2nd magnitude, and so on down to the 5th magnitude, so as to ascertain what law, if any, exists in the distribution of bright and faint stars in the same regions of the heavens? This plan might, I think, be easily carried out for the first three orders of brilliancy, but owing to the large number of stars of the 4th and 5th magnitudes, an extension of the work might give useful employment to several telescopists"

Realising the amount of work this would involve, he later realised that a much more efficient method would be to count stars in photographs. Then, as atlases and catalogues became available to him, he used these as sources of data and carried out various statistical analyses of star distribution. Figure 17 shows such an analysis of naked eye stars in the southern hemisphere carried out in zones of RA and Dec. He found that the average star density in the northern hemisphere was 0.189 stars per square degree and was less than the southern hemisphere. Apart from the region around the Pleiades, the richest northern region was in the vicinity of α and β Cep (0.45 stars per square degree), followed by the region of ε and φ Cas (0.36 stars per square degree) (106). Gore considered some of the dark regions to be truly free of stars. This was surprising as at the time the wide field photographs of E.E. Barnard (1857-1923) were becoming available, and Gore even used some to illustrate his books, which were highly suggestive that the voids were due to obscuring matter.

Gore also noticed that stars often appeared to occur in streams or chains when viewed through a telescope (107), such as the region around 4 and 5 Vul shown in Figure 18a and b. This cluster was first described by the Persian astronomer Al Sufi (903-986) in the 10th century and is nowadays often referred to as Brocchi's Cluster (108) or the Coat Hanger due to its distinctive appearance. The stars are now known not to be physically associated. A modern image of Brocchi's cluster is shown in Figure 18c.

Another problem Gore addressed using the statistical approach that he enjoyed so much was to quantify the brightness of a clear moonless sky (109). He was not the first to do so: Agnes Clerke had obtained a value of 1/80[th] as bright as full moonlight, whereas Monsieur G. L'Hermite had come up with a figure of 1/10[th]. Gore estimated the number of stars in 0.5 magnitude blocks down to magnitude 17.5, a total of 100,198,093 stars and related their combined brightness to the equivalent number of zero magnitude stars (totalling 559 stars). He then added a correction factor to account for the contribution of light by nebulae and the Milky Way, giving a total equivalent to 589 zero magnitude stars. He used a mean of other researchers' data to estimate that full moonlight is equivalent to 46,800 zero magnitude stars, concluding that starlight is equivalent to 1/80[th] of moonlight, which for one hemisphere, all that a single Earth-bound observer can see, is 1/160[th] of moonlight, noting "And this is probably not far from the truth". He also concluded from his analysis that the combined light of all stars fainter than magnitude 6½ "is considerably greater than the light of those above that magnitude, so that if all the stars visible to the naked eye were extinguished we should still have nearly the same amount of starlight". Not content with this treatment, Gore revisited the matter the following year (110), this time basing his calculations on the apparent diameter and brightness of stars as seen from the Earth, but coming up with an identical solution!

Gore's statistical analysis of stars led him to wonder whether in fact the universe is finite or infinite. His found the idea of a finite universe philosophically difficult: "a limit to any space implies a bounding surface, and we cannot conceive of a boundary anywhere, even at a distance far beyond that of the smallest star visible in our largest telescopes, without imagining something beyond that boundary" (111). He then went on point out that if, on the other hand, space were infinite "it would follow that the whole heavens should shine with a uniform light, probably equal to that of the Sun".  This is essentially the problem now commonly known as "Olbers' Paradox" (112). Gore rejected various proposed explanations, such as material between us and the distant stars causing the extinction of starlight. At the time many people thought that light was propagated by a medium known as the "luminiferous ether". The US astronomer Simon Newcomb (1835-1909) suggested in 1878 that this medium was not uniformly distributed, but rather punctuated by "ether voids" that did not allow the light to be further propagated (113). Gore developed this idea in 1888 speculating that the Milky Way was surrounded by such an ether void, or vacuum, which prevented light from external objects from passing though and that light rays striking the edge of the void might be reflected back (Figure 19). Only light originating in our own Milky Way would reach the Earth, either directly from the stars or by reflection from the "verge of the vacuum".  At about the same time Gore was considering voids in the luminiferous ether, 1888, Albert Michelson (1852-1931) and Edward Morley (1838-1923) were conducting their famous experiment, in 1887, which finally disproved the existence of the ether. This was a watershed in physics and it took some time for the idea to be accepted. Gore was still writing about the ether in 1904. However, in 1909 he recognised that the consensus was that the ether did not exist (114).

Gore also considered the possibility of "external universes" beyond our own and concluded that they would be at such "a distance which light would require millions of years to traverse; so that if such universes really exist we should probably know nothing of their existence, as their light would probably not yet have reached the earth". Gore's comments were made in 1888. Lord Kelvin (1824-1907) had addressed Olbers' paradox in one of his 1884 Baltimore Lectures with a similar conclusion: "if all the stars....commenced shining at the same time,....[then] at no instance would light be reaching the Earth from more than an excessively small proportion of all the stars" (115). Whether Gore had heard Kelvin's remarks is not known although it is unlikely as the lecture in which he made the comments was not published until 1901. It is now accepted that the main reason why the sky is dark is that the Universe, and the galaxies it contains, has a finite age and this restricts the amount of light they have emitted (113).

It is also interesting to note that it has been suggested (116) that Gore's idea of "a limited universe, which is isolated by a dark and starless void from any other universes which may exist in the infinity of space" (117) is a precursor to the idea of the "Multiverse" hypotheses which are commonplace in contemporary high energy physics and cosmology. He proposed that there were "systems of higher order" beyond the Milky Way that "Could we speed our flight through space on angel wings beyond the confines of our limited universe...what further creations might not then be revealed to our wandering vision? Systems of a higher order might then be unfolded to our view, compared with which the whole of our visible heavens might appear like a grain of sand on the ocean shore" (118). Of course at the time it was not appreciated that the spiral nebulae revealed in the large telescopes and photographs were separate galaxies, similar to, but beyond, the Milky Way. Nevertheless Gore's vision is pretty close to what we now understand of our galaxy simply being one of billions in the Universe.

**Other interests: astronomical and otherwise**

Lest one think that Gore was solely concerned with the stellar heavens, it should be pointed out that he did occasionally write about other astronomical topics. However, he does not seem to have addressed much observational attention to the solar system, at least in his published work where he often preferred to refer to the work of others, including drawings and other observations by well-known amateurs of the day such as W.F. Denning (119) (1848-1931), T.G. Elger (1838-1897) and N.E. Green (1823-1899). He observed the transit of Venus in 1874 whilst in India (120) and the 1882 transit from western Ireland (121), as well as observing the planet with a telescope on other occasions (122). He recorded a total lunar eclipse in 1891 from Dublin using the naked eye and binoculars (123). He also wrote a review of the Zodiacal Light, referring to his own observations and those of others (124). Whilst in India he observed a variety of nebulae and clusters with his 3 inch refractor (125). Later he published a review of the objects in Messier's catalogue (126), citing historical observations by Messier, Herschel, Smyth and others, as well as the latest perspective revealed by the detailed photographs taken by Isaac Roberts, although he did not include any of his own observations.

Gore was interested in historical aspects of astronomy. On several occasions he wrote about the secular variable stars (127), those stars whose brightness has apparently changed during recorded history. In particular, he compared the relative brightness of stars described by Al Sufi, Ptolemy (ca. 90- ca. 168) and Hipparchus (ca. 190 BC – ca. 120 BC) with their present day brightness, concluding that some stars had brightened and others had faded. For example, based on Al Sufi's writings he concluded that "there can be no doubt that β Leonis has diminished from the first to the second magnitude since the 10[th] century". In ancient times β Leo was recorded as the second brightest star in Leo, not much fainter than α Leo, whereas nowadays it is catalogued as being fainter than γ Leo (128). Whether or not such a change has occurred remains uncertain, but it is not the sort of star that would be expected to undergo brightness variation on this scale of the period of a millennium (129).

Gore was also fascinated by a treatise on astronomy written in Irish in about 1400, commonly known as the *Irish Astronomical Tract*, a copy of which resided in the Royal Irish Academy in Dublin (130). The *Tract* describes the appearance of the celestial sphere and the movements of the stars and planets, as well as covering aspects of physics, geography, meteorology and plant life. The original author is thought to have been an 8[th] century astronomer and astrologer called Messahalah of Alexandria and was probably written in Arabic. The first detailed analysis of the *Tract* was conducted by the Irish cleric Maxwell Close (131) (1822-1903) and published in 1901 (132). Gore felt that "some account of the contents of this valuable tract may prove of interest to the general reader" and published a summary of Close's work, and his own analysis of the English translation, highlighting some of the astronomical points, in *Knowledge and Scientific News* in 1909 (133). It was through Close's and Gore's papers, recognising the value of this early Irish scientific text, that the *Tract* gained more widespread interest and attention.

A strong religious vein ran through much of Gore's writing and he was a man of devout Christian belief. He was also interested in Spiritualism which at the time had a popular following throughout the United Kingdom which included Sir Arthur Conan Doyle (1859-1930). A branch of the Society for Psychical Research was established in Dublin in 1908 (134) and Gore served as its Vice Chairman for a time (135). The Society was originally founded in 1882 by William Fletcher Barrett (1844-1925), Professor of Experimental Physics at the Royal College of Science for Ireland, to undertake psychical research into paranormal phenomena (telepathy, mesmerism), but from the start the Dublin branch was dominated by investigations into the Ouija board, automatic writing, and séances. It was described by one member as "a research group of intelligent, informed and highly placed men and women" and included many well respected members of Dublin society. The Irish poet and playwright W.B. Yeats (1865-1939) was a member of the Society for Psychical Research, but there is no evidence he attended the Dublin branch (136).

Gore's other passion, outside of astronomy, was Chess and he took part in many of the competitions organised through the pages of the *English Mechanic*. For a period, seldom a week passed without him entering.

**Later life**

Although Gore had written his first book, "Southern Stellar Objects for Small Telescopes", whilst in India in 1877, there was a gap until 1888 when he published his next, "Planetary and Stellar Studies" (Figure 20). Like subsequent volumes, this was a popular text, summarising the current state of knowledge of astronomy, with several of the chapters being reproductions or updates on articles that he had published previously elsewhere. Other volumes included "The Scenery of the Heavens" (1890), "Astronomical Lessons" (1890), an observational guidebook: "Star Groups: A student's guide to the constellations" (1891; Figure 21), "An Astronomical Glossary" (1893), "The Visible Universe" (1893; Figure 22), "The Worlds of Space" (1894), "The Stellar Heavens: An introduction to the study of the stars and nebulae" (1903), "Studies in Astronomy" (1904), "Astronomical essays" (1907) "Astronomical Curiosities, Facts and Fallacies" (1910). Many of these books are a delight to read even today, not only for their clarity of prose, but also for the insight they give into the topics of interest in astronomy in the late Victorian and Edwardian era, a time when a great revolution was taking place from the traditional astronomy to the new science of astrophysics. H. G. Wells (1866 – 1946), in his review of Gore's "The Worlds of Space", criticized the author for failing to speculate widely enough on the possible forms that extraterrestrials might take and for not having considered silicone based life forms.

In addition, Gore co-authored "The Concise Knowledge of Astronomy" (1898) with Agnes Mary Clerke (1842-1907) and Alfred Fowler (1868-1940), writing the section on "The Sidereal Heavens". He wrote an introduction to the English translation of Philipp Fauth's (1867-1941) book "The Moon in Modern Astronomy" (1907) and he translated Camille Flammarion's (1842-1925) work "Popular Astronomy" (1894) from the French. This translation went through several editions in Britain and the USA. Flammarion was evidently delighted by Gore's proposal to translate his book, replying on 17 April 1893: (6)

"Nothing can be more welcome to me than your gracious proposal. I have all your works in my library and I esteem them very much. We pursue the same end – progress through light.....I am very glad that this opportunity has arisen to enable me to express to you deep regard which I have long had for your own splendid volumes"

However, none of the books of which he was the sole author ran to a second edition, in contrast to those of his compatriot Sir Robert Ball. This was partly because the contents quickly became outdated as scientific knowledge increased and partly because Gore wrote so many in quick succession that much of the material was repeated, with updates, in subsequent volumes. It cannot be denied that Ball's engaging style of writing might have had greater appeal to the general reader.

Outside of astronomy, in 1874 Gore also published a "Glossary of Fossil Mammalia for the Use of Students of Palaeontology" (137).

Gore's eyesight began to fail after about 1900, restricting his astronomical activities, which in any case had largely been superseded by his interest in theoretical astronomy. No longer able to read easily, in 1909 he donated his library to the Royal Irish Academy. On the evening of Monday 18 July 1910, Gore was crossing Grafton Street, Dublin, when he failed to see a jaunting car (138) approaching. He was struck on the head and he was thrown beneath the wheels. A Dublin doctor was passing in a motor car and stopped to give assistance, but he was pronounced dead within a few minutes. His body was brought to Mercer's Hospital and was later buried in Mount Jerome Cemetery in Dublin (139). He was 65 years old. The driver of the jaunting car was indicted for causing Gore's death, but was subsequently acquitted. The judge said "There was no doubt about it, that any person crossing the streets of Dublin took his life in his hands" (6). Gore's failing eyesight may well have been a contributory factor. His last letter to the *English Mechanic* appeared in the edition of 1 July 1910 (140).

### Reflections and perspectives

It spite of detailed obituaries appearing elsewhere, Gore's JBAA obituary only ran to a few lines. He is largely unknown in BAA circles today and his contribution to the fledgling organisation is perhaps underestimated. Perhaps one reason for this is that his role as VSS Director was not generally viewed as particularly successful, although he was working within the mandate for a section Director. For example, we have seen that he did not encourage cooperation between observers and that towards the end of his Directorship membership was sliding. Howard Kelly's comment in his note on the first 50 years of the Section, which forms part of the official history of the BAA, is as telling as it is dismissive (141). He spends a single sentence on Gore's Directorship, then goes on to say that "It was not until 1900 that the activities of the members became adequately organised, when, under the Directorship of Markwick, observational work was established on fully co-operative lines".  It is true that Markwick essentially re-started the Section, bringing his energy and enthusiasm to bear, and organised it along the lines on which it continues today, as one of the most active observing sections in the BAA. One could say that Gore's contributions to the BAA have been eclipsed for more than a century by Markwick's long shadow. But it is always dangerous to judge a man by a single contribution and in Gore's case this is particularly unfair, since as we have seen he made so many contributions in different areas.

Gore ranks alongside other notable British variable star observers of the second half of the 19[th] century - Norman Pogson (1829-1891), George Knott and Joseph Baxendell – not only as an observer, but also through his discovery of new variables at a time that relatively few were known. He brought the latest advances in variable star astronomy to many people through his writings and he showed that it was a branch of observational astronomy well suited to amateurs with modest means who wished to perform scientifically useful work.

Gore was also at the cutting edge of binary star science through his computation work on orbits and elements. This culminated in his work on systems such as Sirius, where the companion was thought possibly to be a dark body. Through logical reasoning, and basic calculations he showed that the companions are in fact self-luminous. He was also the first person to show that these companions, now known as white dwarfs, might have an extremely high density. His calculation of the density of Sirius B, although way off by modern standards and which he ultimately rejected as being beyond the realms of possibility, is the first estimate of the density of a white dwarf. This probably stands as his greatest single contribution to astronomy. Apart from this, he was also among the first to realise that stars exist in a great range of size from "miniature stars" (dwarf stars as we would call them today) to giants.

By contrast, his thoughts on the structure of the Universe soon became outdated as new concepts in physics and astronomy emerged, such as how light is transmitted – overturning the need for a luminiferous ether - and the fact that our Milky Way is merely one among billions of galaxies, or "island universes". However, his work on the size of the universe addressed fundamental problems such as why the night sky is dark (Olbers' paradox) was at the cutting edge at the time.

Whilst he may not have been quite the equal of the great astronomy author of the time, Sir Robert Ball, Gore's books helped to feed the insatiable public appetite that existed at the time for knowledge about developments in science. Not only did he bring to his readers' attention the latest theories on the structure of the Universe, but he also touched on the developments at the atomic scale that were occurring in physics. Thus he covered the range of immensity and minuteness of the physical world. Whilst some of what he wrote in his books did not withstand the test of time, as the new science of astrophysics blossomed, they can still be read with profit today for the insight they give to an era of tremendous scientific progress. Gore's contribution to advancing popular astronomy was also in evidence through his support for the BAA from its inception and during its first two decades.

This paper, written slightly over a century after his death, has focussed on Gore's work and achievements. So what of Gore the man? Gore's close friend, Monck, described him as being "of a quiet retiring disposition" (4); he was also described as a grave man, with few friends, but very much liked by all who knew him (3), noted for his quiet wisdom and gracious courtesy (8). The astronomical author Hector Macpherson (1851-1924) corresponded frequently with Gore (142), although they never met, and described him as "kindness itself and an attentive and interesting correspondent.....He never held any official position as an astronomer and received no honorary degrees; yet his astronomical work takes high rank among the observers of his day and generation. His life....is a brilliant example of what an amateur may accomplish in astronomy" (5)

In recognition of Gore's contributions to astronomy, on 22 January 2009 the International Astronomical Union named a 9.4 km crater on moon after him (143).

One of the aims of this paper has been to make Gore's work more widely known to members of the BAA. I wanted to put into perspective the full extent of his pioneering contributions to astronomy (some, it could be argued, were way ahead of their time), in the fields of variable stars, binary stars, stellar size, the structure of the Universe, the popularisation of astronomy and the development of the BAA. He certainly deserves to be better known.

## Acknowledgements

I am most grateful for the assistance I have received from a large number of people during the research for this paper. Jay Holberg (Lunar and Planetary Laboratory, University of Arizona, USA) generously provided copies of his papers and presentations in connexion with his research on Gore. Eric Hutton produced an extensive extract of items from the *English Mechanic* relating to Gore. Ian Elliott provided information which he had used to prepare Gore's entry for *The Biographical Encyclopaedia of Astronomers* and he pointed me in the direction of the article about Gore by Charles Mollan in his series written for the Royal Dublin Society on science in Ireland, '*It's Part of What We Are'.* Pat Corvan provided insights in Irish astronomy and alerted me to the archive at Armagh Observatory which contains papers related to Gore and I thank John McFarland (Curator of Archives, Armagh Observatory) for providing detailed information on these papers. Bernadette Cunningham (Deputy Librarian Royal Irish Academy) provided information about Gore's links to the RIA. Randall Rosenfeld (Royal Astronomical Society of Canada) provided details of Gore's membership of the Astronomical and Philosophical Society of Toronto. Jerry Lodriguss kindly gave permission to his image of the Cygnus-Cepheus region of the Milky Way (Figure 3) and Alan Tough allowed me to use his image of Collinder 399 (Figure 18c). Professor James Overduin (Townson University, USA) gave permission to use the diagram of Gore's ether voids which appeared in reference (113). Mike Saladyga (AAVSO) provided copies from the AAVSO archives of Gore's observations of W Cygni that Gore had sent to Yendell. Jean Poyner kindly tracked down and obtained copies of Gore's papers published in the Proceedings of the Royal Irish Academy. Richard Baum provided much encouragement and inspiration to pursue this research. Peter Hingley (RAS Librarian) looked after me on numerous visits to the RAS library and provided details of Gore's RAS Fellowship application and his portrait shown in Figure 1. John Toone provided the portrait of George Knott. To all these people I am hugely indebted. I would also like to thank the referees,

## Address

"Pemberton", School Lane, Bunbury, Tarporley, Cheshire, CW6 9NR, UK

13. Elizabeth Sophia Rebecca Gore (1850-1921).

14. As well as the Diploma, Gore was awarded an additional Certificate for Practical Engineering by the School of Engineering at Trinity College. Both were awarded on 12 December 1865. Archives of Armagh Observatory: location code B9.7.4, object M 187.G17.

15. Newspaper article in the archives of Armagh Observatory: location code B9.7.4, object M 187.G34.

16. The Sirhind Canal was opened in 1882 and consists of an extensive canal system that irrigates more than 2,000 square miles (5,200 square km) of farmland. The canal begins at Ropar headworks near Ropar City in Rupnagar district of Punjab.

17. A.P. FitzGerald, in his IrAJ biography of Gore, says that he was "an outspoken young man", implying that this contributed towards the tension with his boss.

18. Letters in the archive of Armagh Observatory indicate that he contacted Robert S. Ball and Edmund Neison at Natal Observatory about employment prospects. He also investigated openings in surveying. He hadn't ruled out a return to India, for in 1884 he corresponded with the Public Works Department about jobs. There are also several health and character references, which he presumably used to support job applications.

19. The MNRAS obituary makes this suggestion, which was repeated in A.P. Fitzgerald's biography of Gore.

20. Gore was elected Member of the RIA on 12 April 1875; Doberck was elected in January 1876.

21. Gore used this telescope for many of his double star observations reported in the English Mechanic and his resolution tests showed it to be of excellent quality, e.g. English Mechanic, 478, item 7522 (1874).

22. Gore J.E., Southern Stellar Objects for Small Telescopes, between the equator and 55 degrees south declination, publ. Lodiana Mission Press (1877).

23. The following is a quote from a personal letter written by Gore to W.S. Franks and subsequently published in a letter to the English Mechanic by Franks: Franks W.S., English Mechanic, 1121, item 26247 (1886).

24. The present author lived in the United Arab Emirates for two years. Whilst the visibility from the desert in the UAE and Oman could be excellent during the winter, during the summer I experienced similar conditions to those reported by Gore. The elevated temperatures combined with fine dust in the atmosphere made observing challenging.

25. Gore J.E., English Mechanic, 395, item 5072 (1872). Gore wrote this letter on 7 September at Kussowlee, India. He also describes the observation of this "Cygnus coal sack" in his later books, e.g. Chapter 15 in Gore J.E., Planetary & Stellar Studies, publ. Roper and Drowley (1888) and in Gore J.E., AReg, 11, 50 (1873).

26. Gore J.E., AReg, 11, 50 (1873).

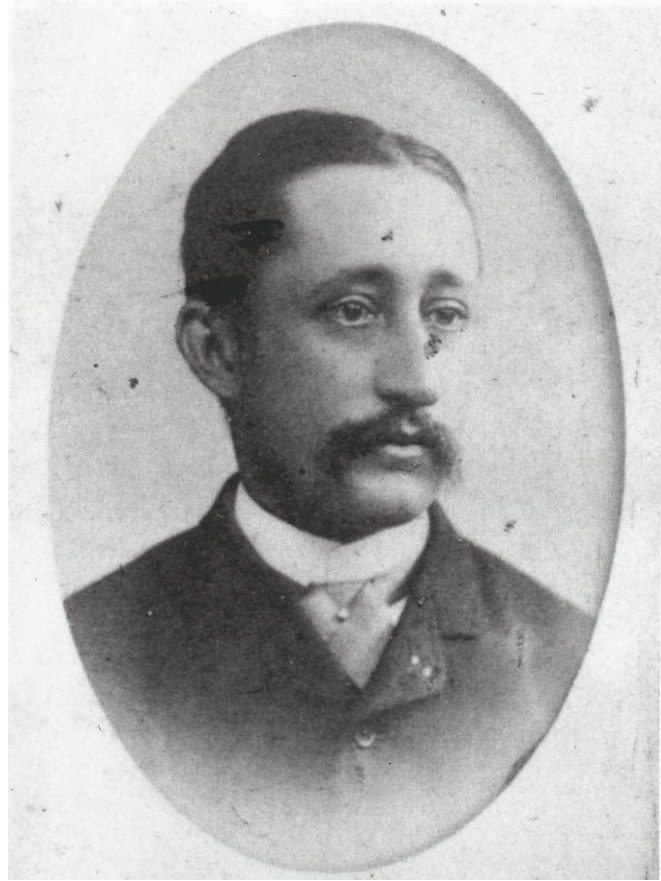

Figure 1: John Ellard Gore

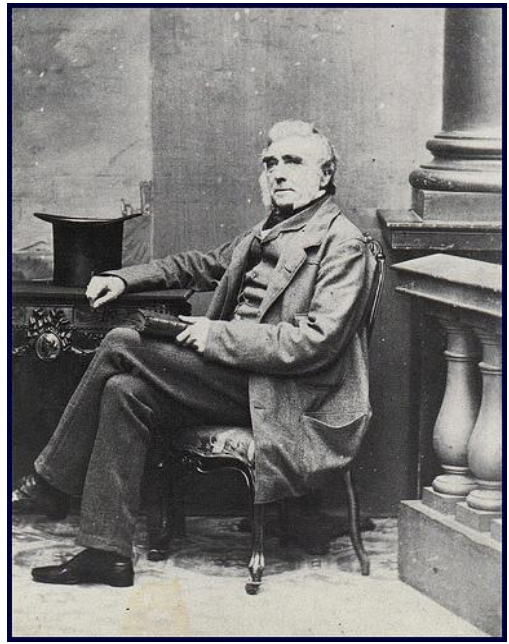

Figure 2: Edward Joshua Cooper (1798-1863)

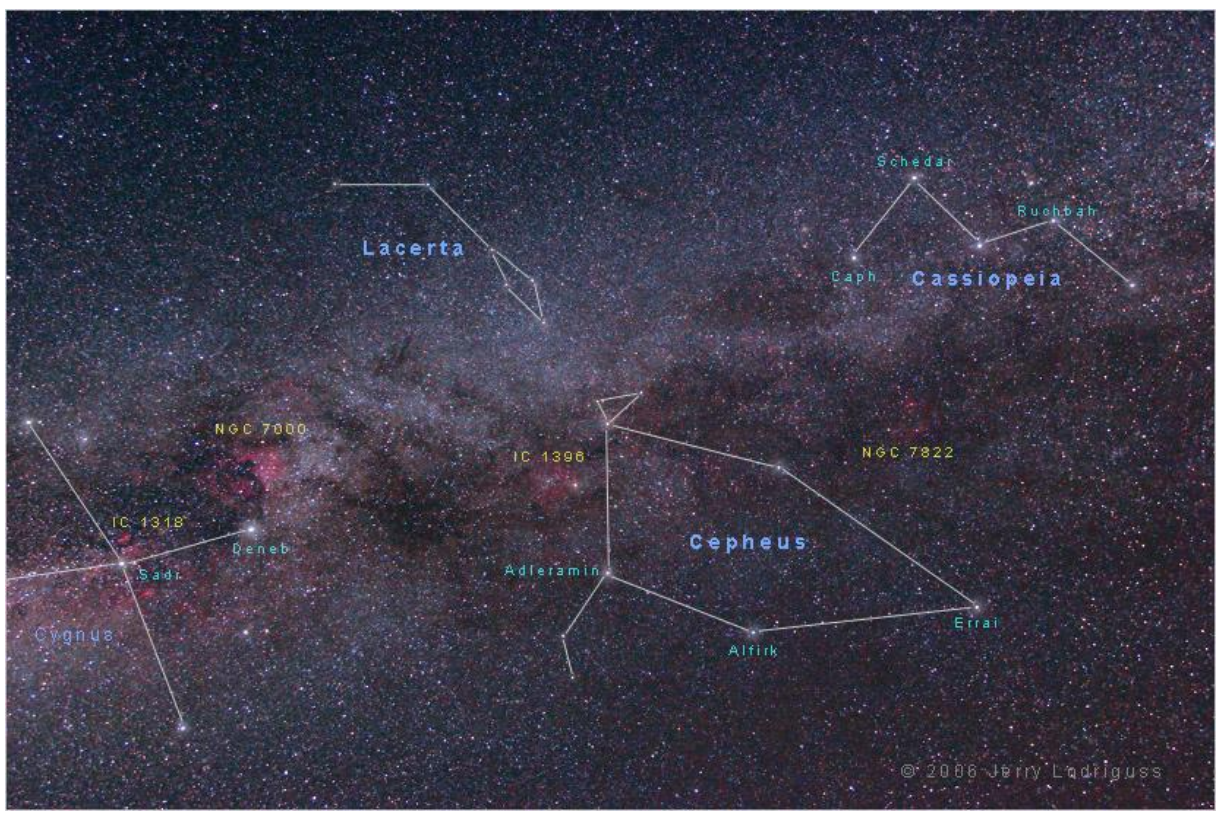

Figure 3: The Cygnus-Cepheus area of the Milky Way. Gore's "remarkable vacuity" can be seen between NGC7000 (the North American Nebula) and IC 1396. Imaged by Jerry Lodriguss on 3 September 2005, using a Canon EOS 20Da DSLR camera, a Canon 16 - 35mm zoom lens working at 20mm, and an 800 ISO setting. Exposure: composite of 5 frames, each 5 minutes exposure, 25 minutes total exposure

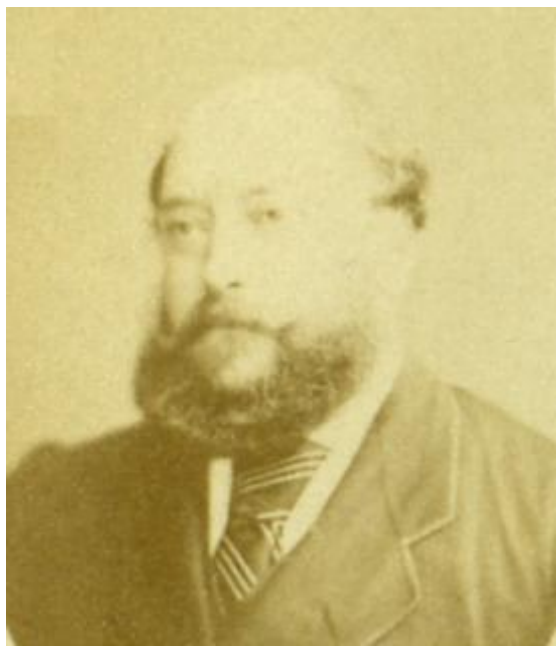

Figure 4: George Knott (1835-1894)

Figure 5: Gore's list of maxima and minima of W Cyg which he sent to Paul Yendell (AAVSO archives)

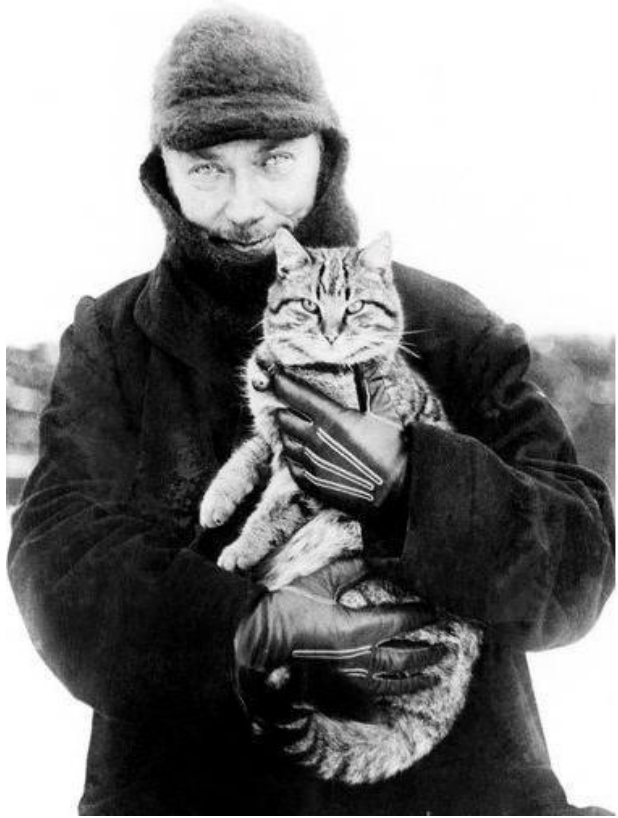

Figure 6: T.H.E.C. Espin (1858-1934)

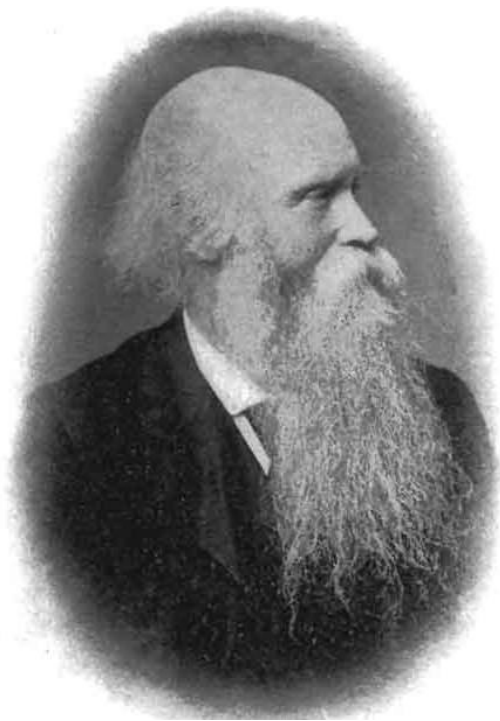

Figure 7: Ralph Copeland (1837-1905)

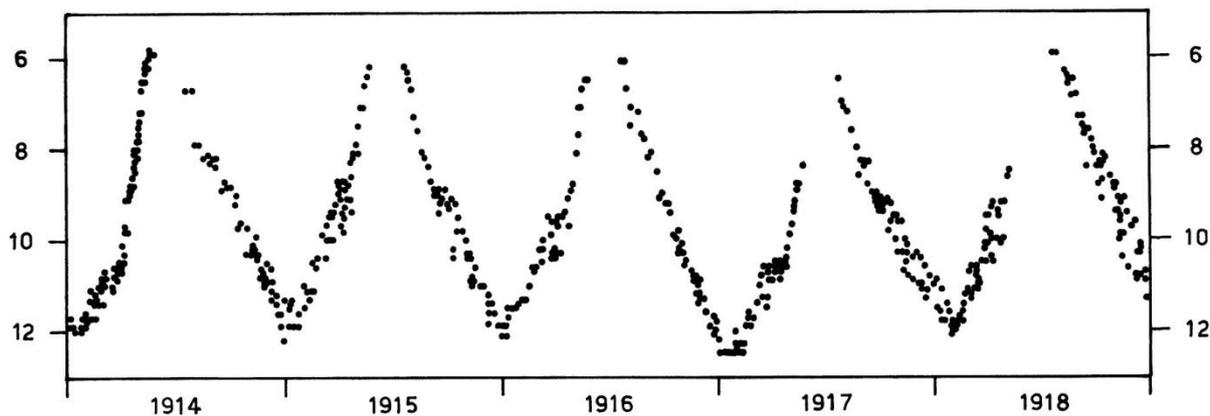

Figure 8: Light curve of U Orionis (46)
(John Toone)

Figure 9: Gore's RAS Fellowship proposal form.
Proposed: 11 January 1878. Elected: 8 March 1878
(courtesy Peter Hingley, RAS)

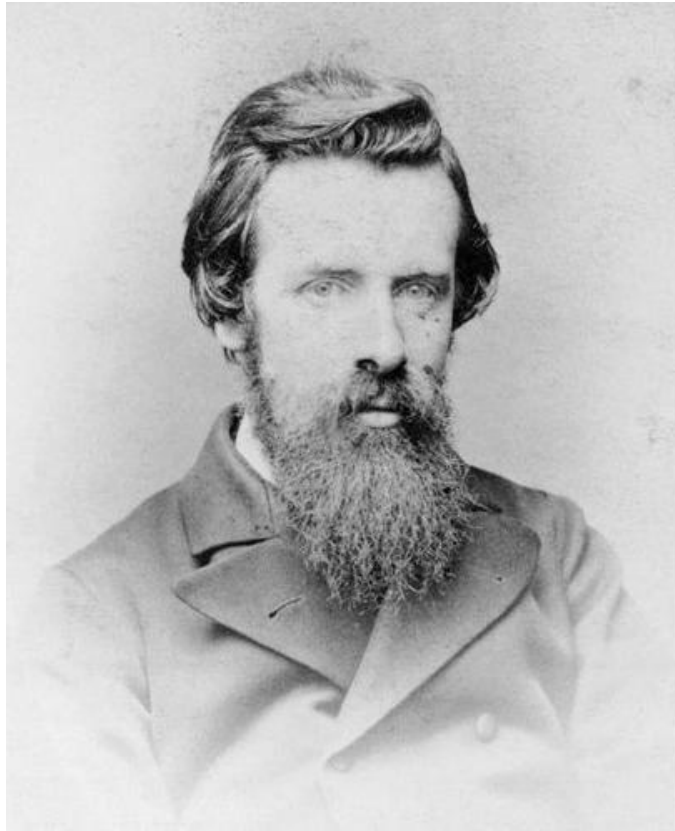

Figure 10: John Browning (1835-1925)

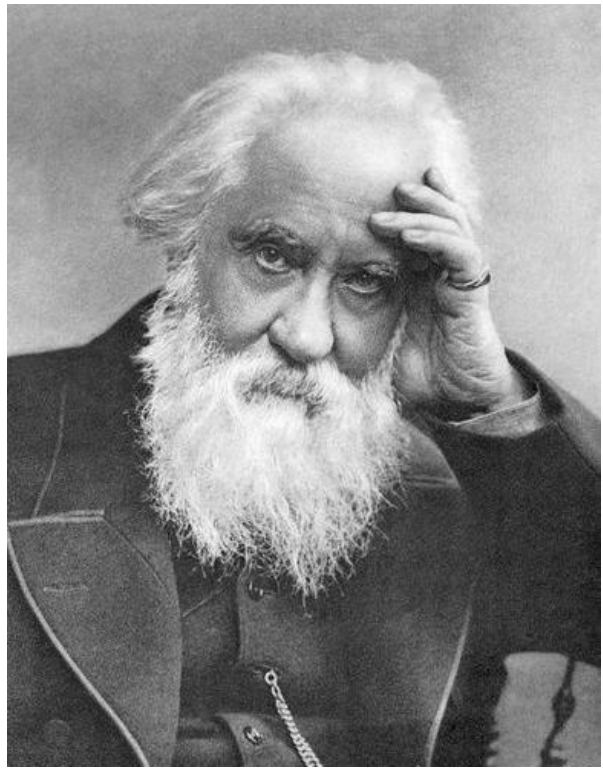

Figure 11: William Huggins (1824 – 1910)

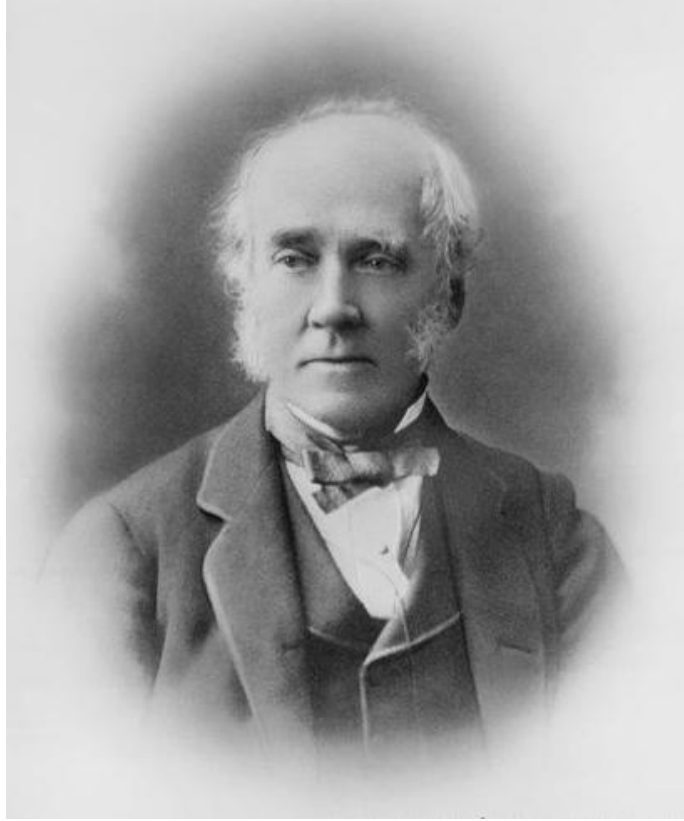

Figure 12: William Lassell (1799-1880)

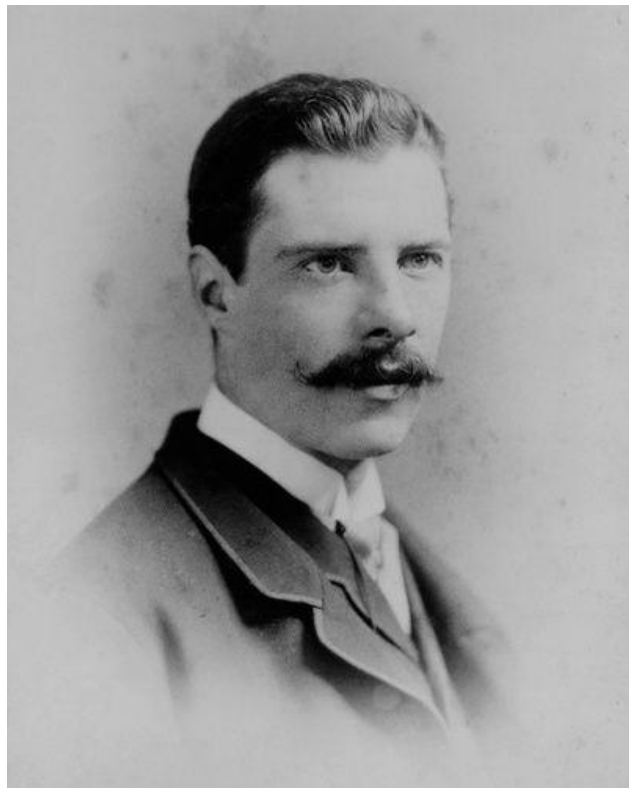

Figure 13: Edmund Neison (1849-1940)

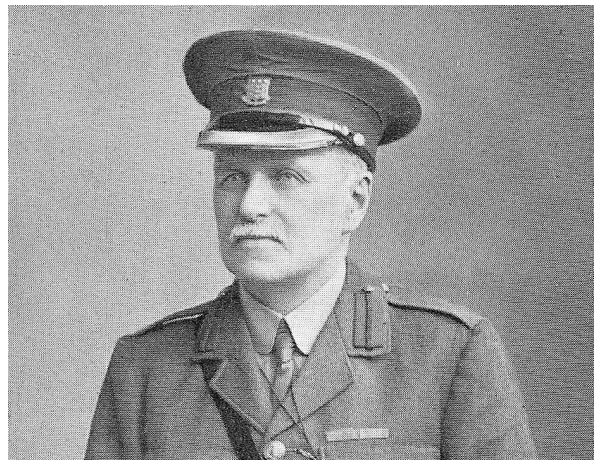

Figure 14: Col. E.E. Markwick, CB, CBE, FRAS (1853 – 1925), Gore's successor as BAA VSS Director

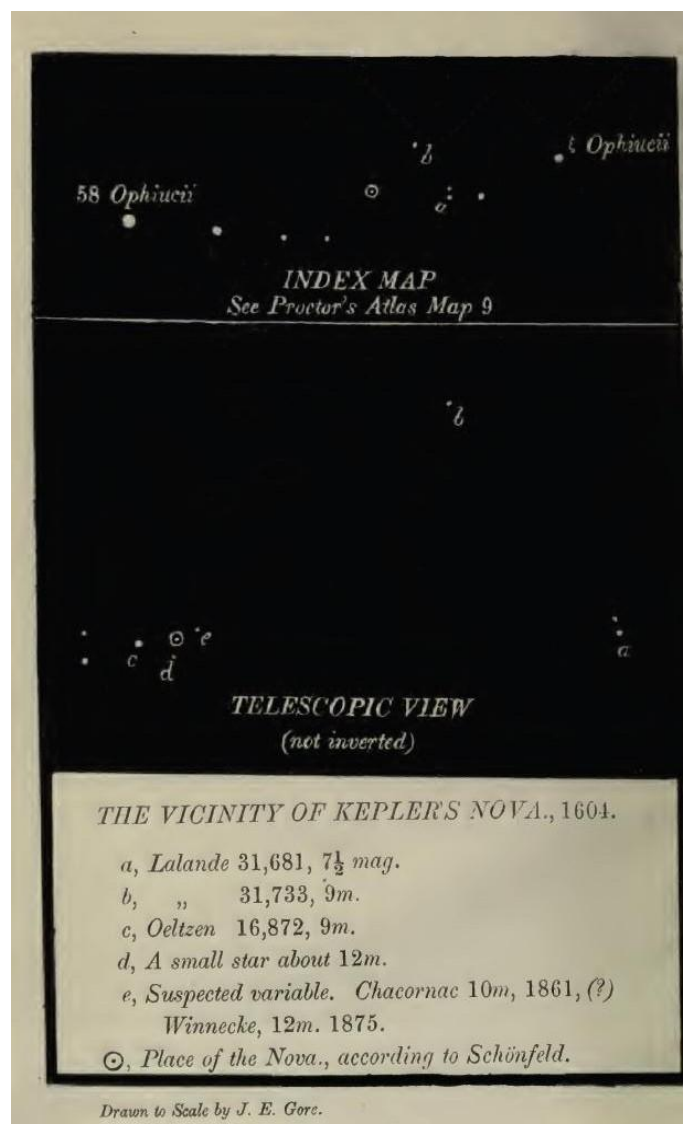

Figure 15: Gore's chart of the position of the supernova of 1604 from his book *Southern Stellar Objects for Small Telescopes*

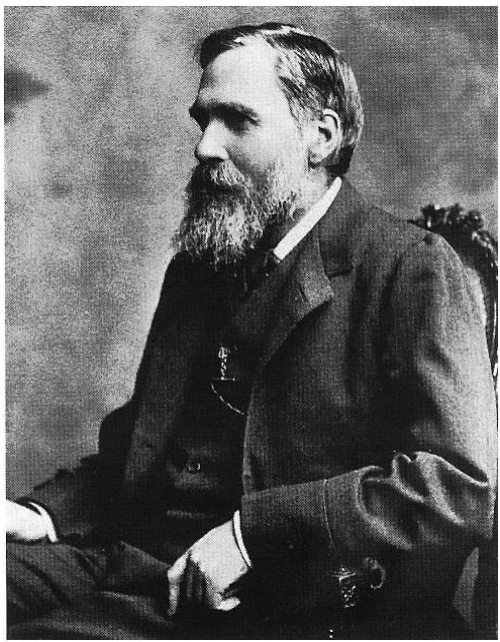

Figure 16: William Henry Stanley Monck (1839-1915)

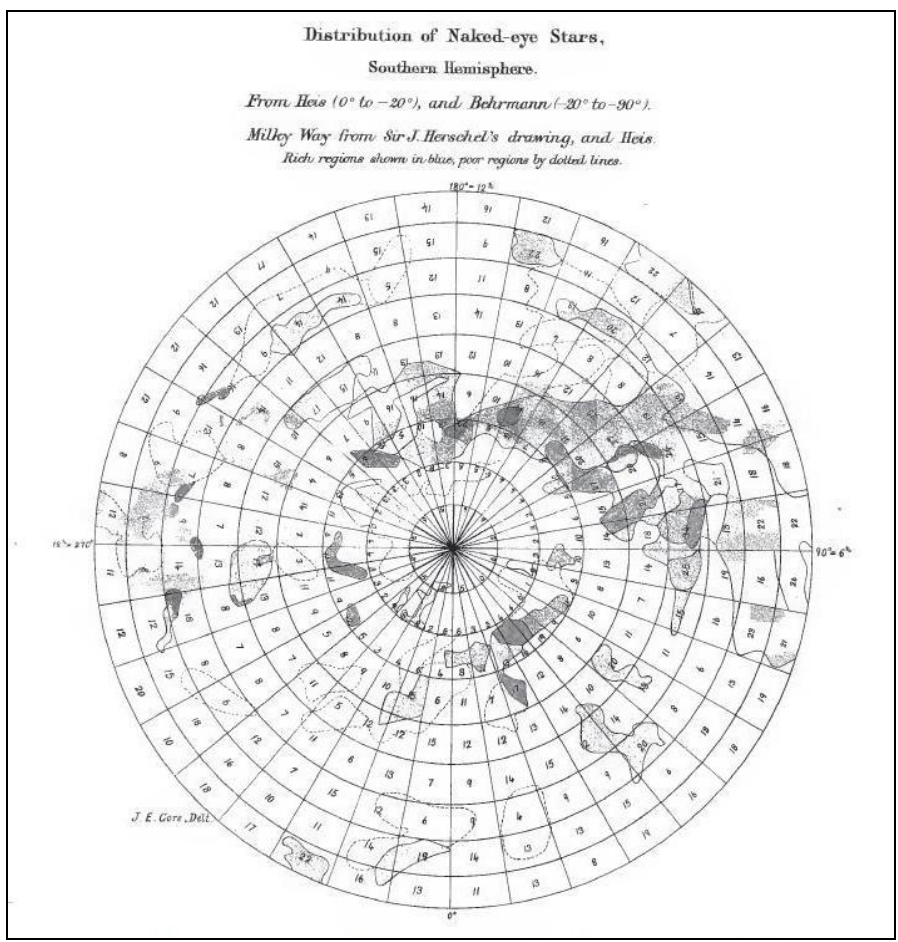

Figure 17: Distribution of naked eye stars in the southern hemisphere. Gore's drawing shows the Milky Way (shaded areas) and is measurements of the number of stars in zones of RA and Dec. From reference (106).

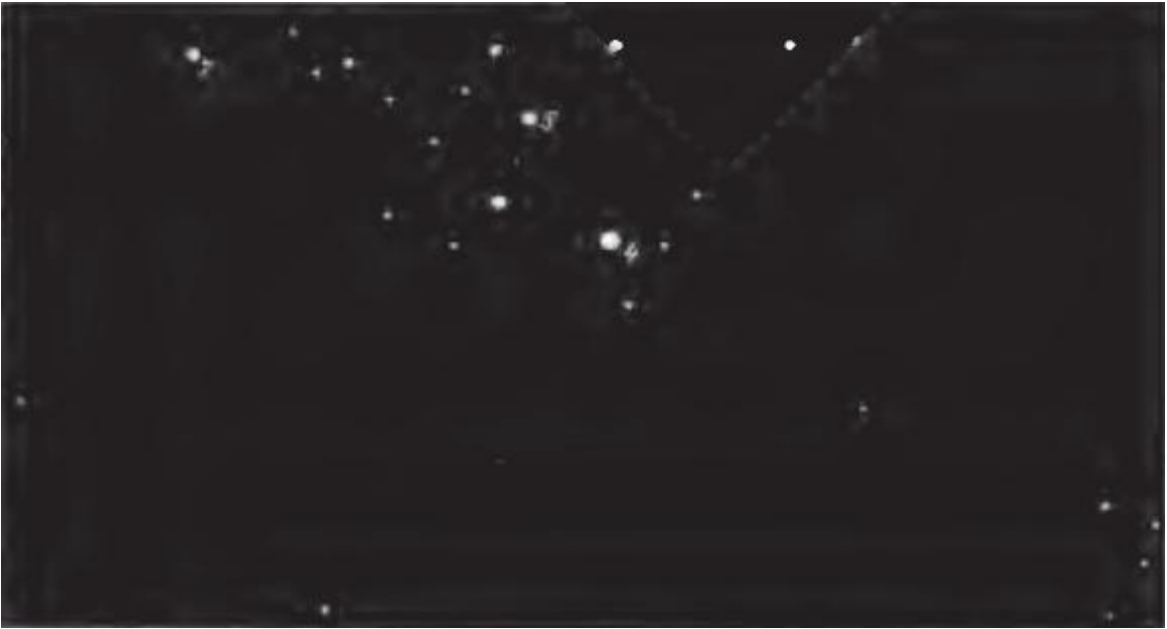

(a)

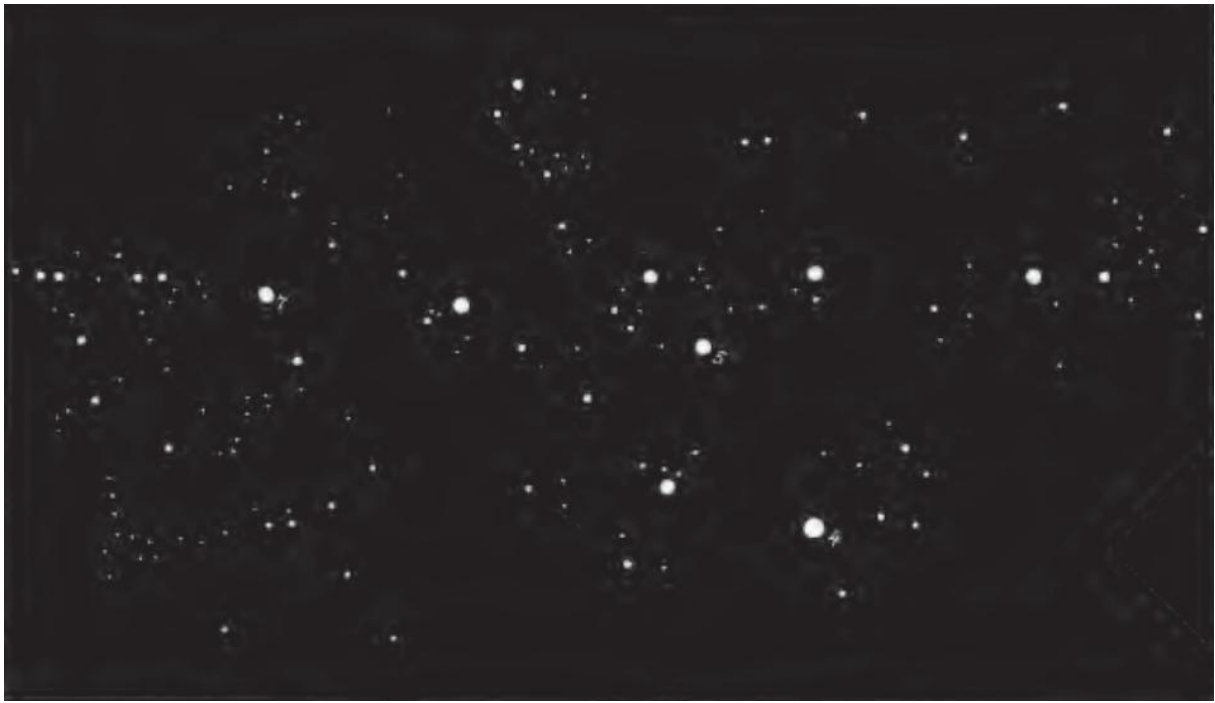

(b)

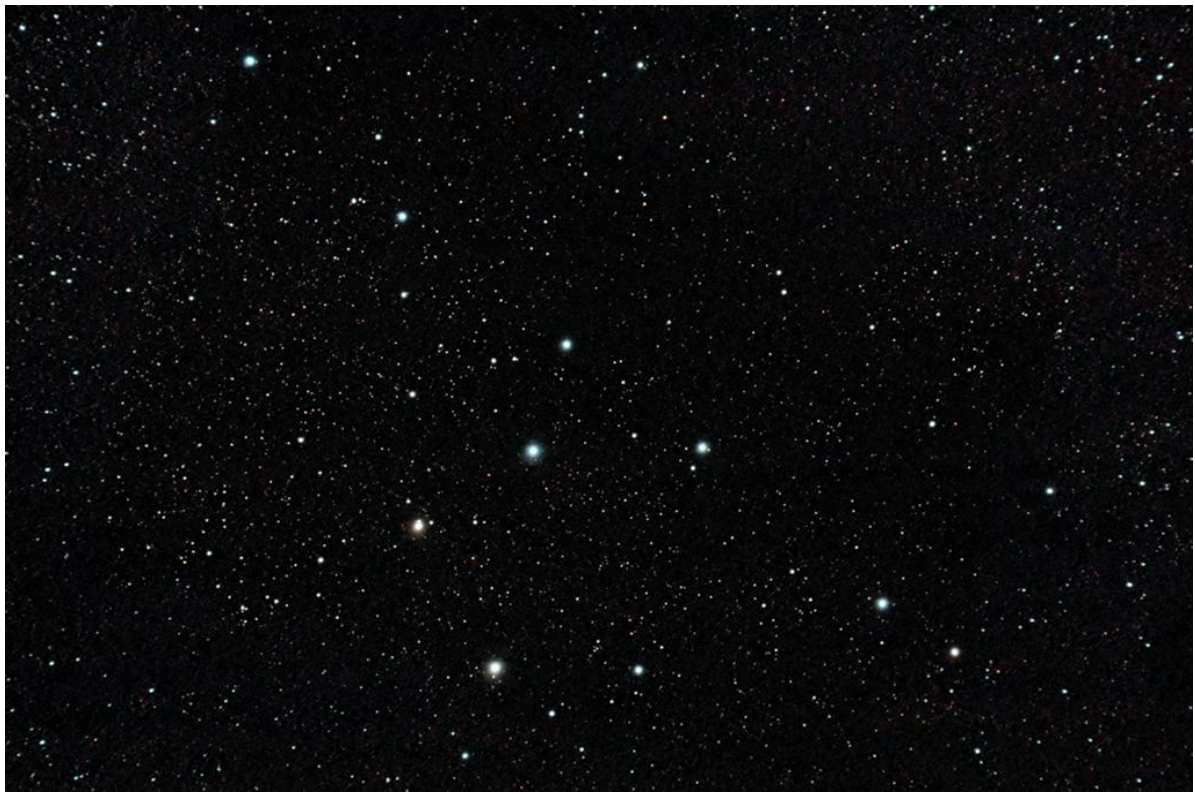

(c)

Figure 18: "Star streams" in Vulpecula, the cluster now known as Collinder 399 or Brocchi's Cluster.

(a) Drawing made by Gore with a 2 inch refractor in October 1889, from reference (107)
(b) Drawing made by T.H.E.C. Espin with his 17¼ inch reflector x70, showing stars to magnitude 10½, 5 November 1889, from reference (107)
(c) A modern image of taken by Alan C. Tough on 18 September 2009. Equipment: Sky-Watcher ED80 refractor, HEQ5 mount, Canon EOS 40D at prime focus. Exposure: 3 x 5 minutes @ f/7.5, ISO-800

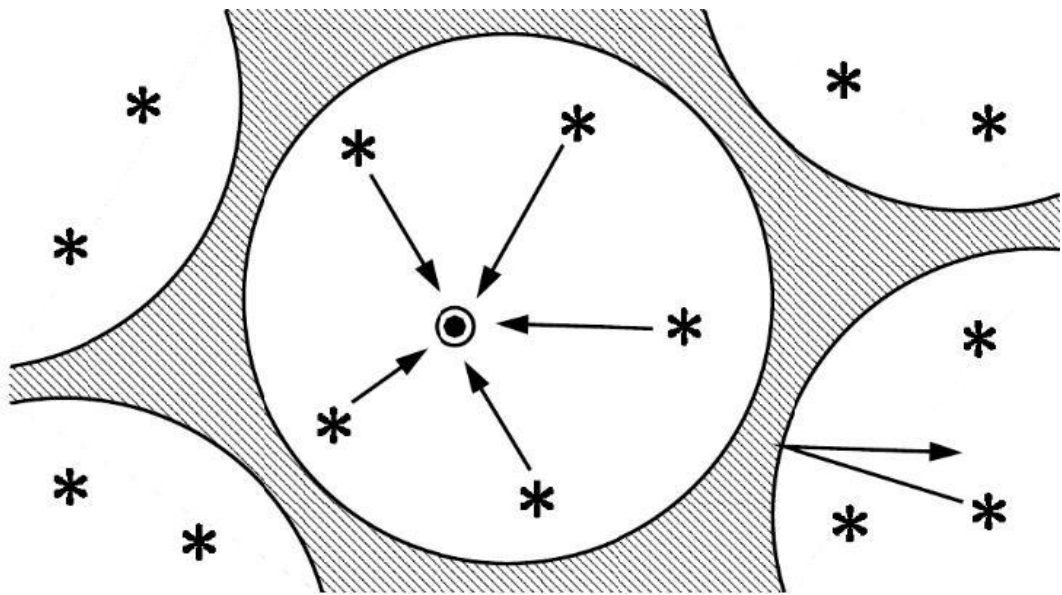

Figure 19: Ether voids. One idea that Gore presented to explain why the sky is dark at night is due to ether voids which prevent light from stars in distant parts of the universe passing, instead the light rays may be reflected back (Image from reference (113), used with permission.

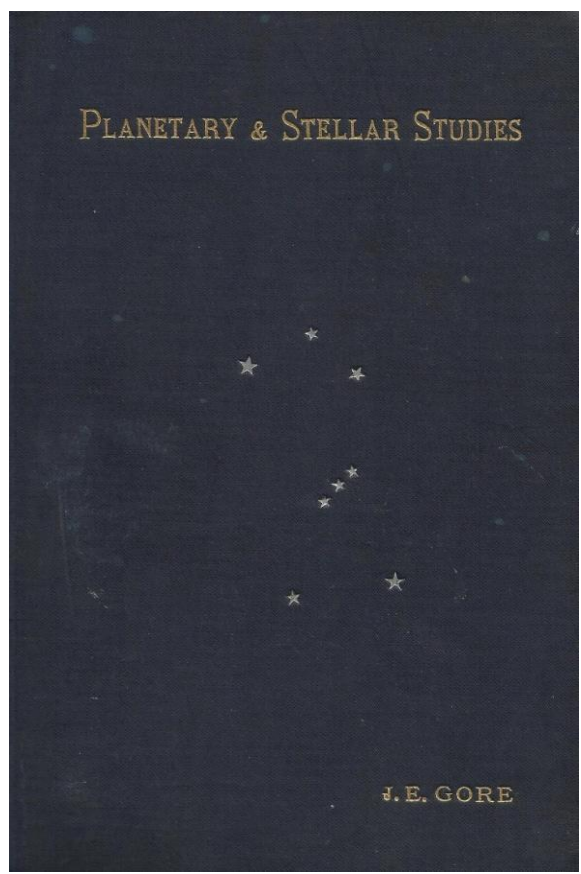

Figure 20: Cover of Gore's *Planetary & Stellar Studies* (1888)

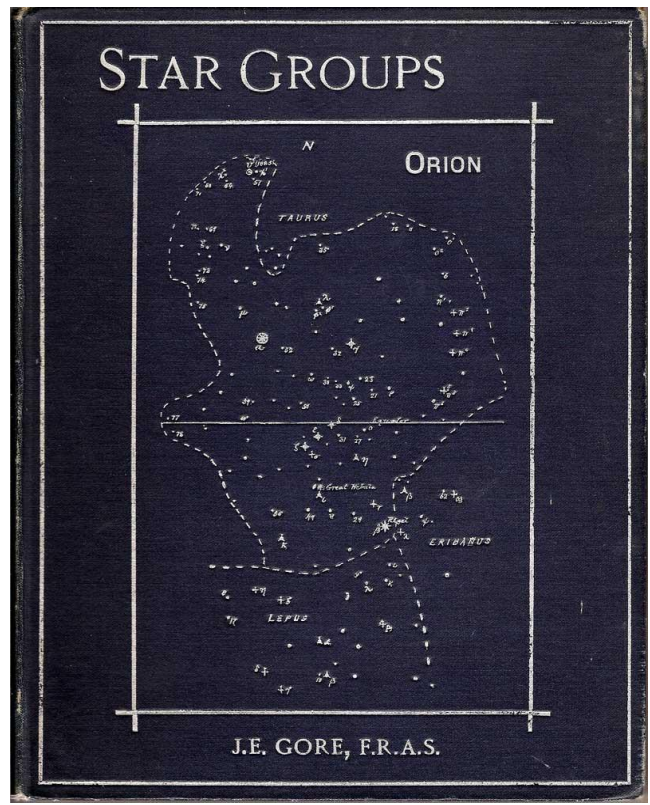

Figure 21: Cover of Gore's *Star Groups: A student's guide to the constellations* (1891)

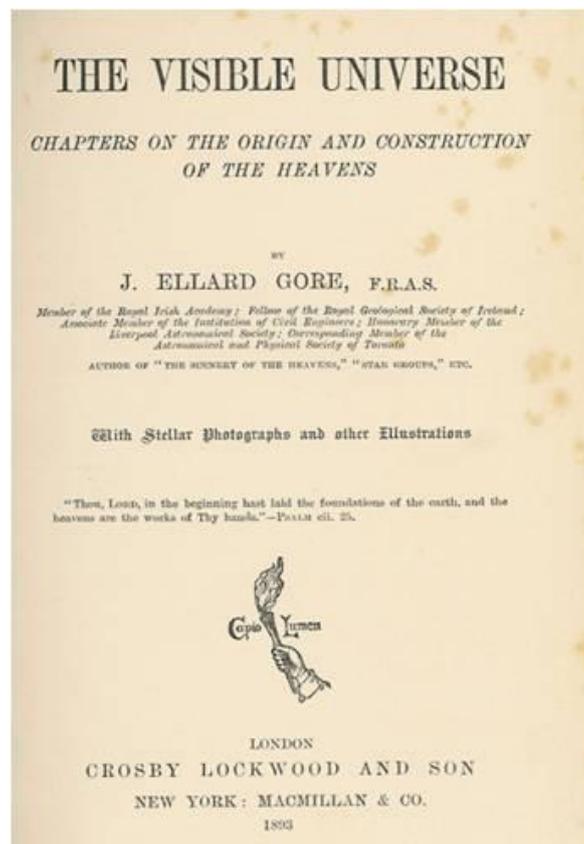

Figure 22: Title page of Gore's "The Visible Universe" (1893)

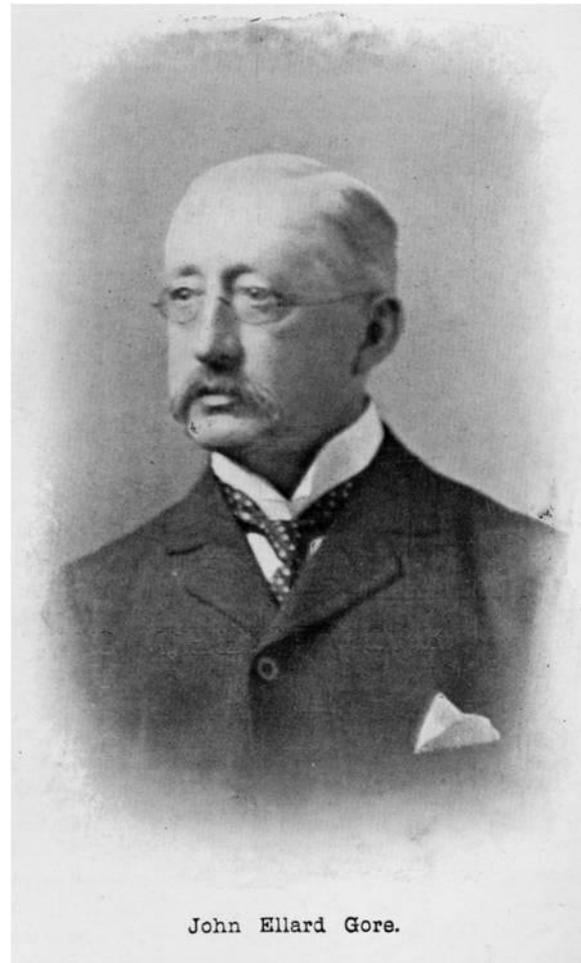

John Ellard Gore.

Figure 23: Gore in later life

(from reference (5))